\documentclass[twocolumn,showpacs,preprintnumbers,amsmath,amssymb]{revtex4}
\usepackage{graphicx}% Include figure files
\usepackage{dcolumn}% Align table columns on decimal point
\usepackage{bm}% bold math

\begin{document}

\preprint{APS/123-QED}

\title{Nonclassical rotational behavior at the vicinity of the $\lambda $ point}% Force line breaks with \\

\author{Shun-ichiro Koh}
 
 \email{koh@kochi-u.ac.jp}

\affiliation{ Physics Division, Faculty of Education, Kochi University  \\
        Akebono-cho, 2-5-1, Kochi, 780, Japan 
}%

\date{\today}% It is always \today, today,

\begin{abstract}
The rotational property of a quantum liquid at the vicinity of the $\lambda$
point $T_{\lambda}$ is examined. In a liquid helium 4 just above $T_{\lambda}$,  
 under the strong influence of Bose statistics, 
 the coherent many-body wave function grows to an 
 intermediate size between a macroscopic and a microscopic one, which is 
 of a different nature from the thermal fluctuations.  
 It must reflect in the rotational properties such as the 
 moment of inertia.  Beginning with the bosons 
without the condensate, we make a perturbation calculation  of its 
susceptibility with respect to the repulsive interaction, and examine how, with 
decreasing temperature, the growth of the coherent wave 
function gradually changes the rotational behavior of a liquid:
 The moment of inertia  slightly decreases just above $T_{\lambda}$. This 
 means that at the vicinity of $T_{\lambda}$, the 
 mechanical superfluid density does not always agree with the thermodynamical one.
 We compare the result to the experiment by Hess and Fairbank.  A new 
 interpretation of the shear viscosity just above $T_{\lambda}$ is given 
 from this viewpoint. 
\end{abstract}

\maketitle

\section{Introduction}

 A natural way to discuss  superfluidity in a confined system is to focus on its 
rotational properties.  When a liquid helium 4 is rotated at a temperature far
above the $\lambda$ point $T_{\lambda}$, it makes a rigid-body 
rotation with a uniform vorticity  $rot \mbox{\boldmath $v$}\ne 0$ owing 
 to its viscosity.  The angular momentum around z-axis has a form of 
$L_z^{cl}=I_z^{cl}\Omega$, where $I_z^{cl}$ is a classical moment of inertia 
and $\Omega$ is a rotational velocity of a container. Figure.1 
schematically shows the  $\Omega$ dependence of the angular momentum. 
When it is cooled to a certain temperature below $T_{\lambda}$, it 
shows two different behaviors according to the value of  
$\Omega$. Below the critical velocity 
$\Omega _c$, it abruptly stops  rotating just as the system passes the $\lambda$ 
point, and  $rot \mbox{\boldmath $v$}=0$ is satisfied over the whole volume 
of the liquid. On the other hand, under a faster rotation than $\Omega _c$, 
the  uniform vorticity abruptly concentrates to certain points, forming 
vortex lines and leaving other areas to satisfy $rot \mbox{\boldmath $v$}=0$. 
 These phenomena instilled us with the notion that 
superfluidity abruptly appears at the $\lambda$ point. 

The basis of our phenomenological understanding is the two-fluid model, 
 the foundation of which at $T \ll T_{\lambda}$ has been established by microscopic 
theories. The basic assumption of the two-fluid model is that it completely 
separates the system into the normal and  
superfluid part from the beginning, and assumes that the 
latter abruptly emerges at $T_{\lambda}$ \cite {var}. Furthermore, it assumes 
that the superfluid density defined in the mechanical properties 
completely agrees with that defined in thermodynamics.

Here we must make a clear definition of superfluidity. Just above 
$T_{\lambda}$, anomalies in thermodynamical or mechanical 
constants of a liquid, such as the $\lambda$ shape of specific heat or 
the softening of sound propagation, 
has been observed; These anomalies are normally considered to arise 
from thermal fluctuations giving rise to the short-lived randomly oriented coherent 
wave functions.  For superfluidity, however, 
 the long-lived collective motions of particles with a stable 
 specific direction of motion are necessary. London 
stressed that superfluidity is  not merely the absence of 
viscosity, but the occurrence of $rot  \mbox{\boldmath $v$}=0$, 
 and proposed an experiment to confirm this point \cite {lod}, which was 
 later performed by Hess and Fairbank \cite{hes}. 
 This means that the complete disappearance of 
shear viscosity is attributed to $rot \mbox{\boldmath $v$}=0$ over the 
whole volume of a liquid. Hence, the thermal fluctuations, being a 
collection of independently growing and decaying wave functions, do not 
lead to superfluidity. 
\begin{figure}
\includegraphics [scale=0.4]{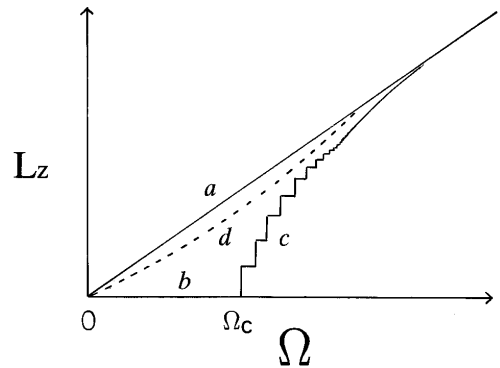}
\caption{\label{fig:epsart}
The schematic $\Omega$-dependence of the angular momentum $L_z$. A solid 
 straight line {\it a \/} is $L_z=I_{z}^{cl}\Omega$ at 
 $T \gg T_{\lambda}$, and a horizontal line {\it b \/} and a series of 
 steps {\it c \/} is $L_z(\Omega)$ at $T<T_{\lambda}$. A 
 dotted curve {\it d \/} is a subject of this paper, a size of which is 
 exaggerated for clarity.}
\end{figure}
(The relaxation time of thermal fluctuations is 
far shorter than the characteristic time of the macroscopic rotational 
experiments. Hence, they decay long before they affect the moment of inertia.)

 This paper will examine the foundation of the two-fluid model at the 
 vicinity of the lamda point. Just above $T_{\lambda}$, particles experience the strong influence of 
Bose statistics, and therefore, the thermal-equilibrium  coherent many-body wave function 
satisfying  $rot \mbox{\boldmath $v$}=0$ grows to a large but not yet 
macroscopic size. In contrast to thermal fluctuations, these long-lived wave functions 
have a possibility of affecting a  mechanical property of the whole
liquid, such as the moment of inertia. Specifically, they may
slightly reduce the moment of inertia  $I_z$ just above $T_{\lambda}$, a
possible $L_z(\Omega)$ of which is schematically illustrated by a dotted 
line {\it d \/} in Fig.1.  This is qualitatively different from 
precursory phenomena owing to thermal 
fluctuations in that it requires a stable specific direction of motion. 
If it is true, it allows us to redefine the superfluid density in the mechanical 
phenomena. This means that at the vicinity of 
$T_{\lambda}$, the  mechanical superfluid density 
 does not always agree with the thermodynamical superfluid density.

The rotational properties of a liquid helium 4 has been subjected to 
considerable experimental and theoretical studies \cite{don}. These studies, 
however, mainly focus on the  dynamics of the quantized vortices in the 
superfluid phase in situations where the rotational velocity is not so 
small that the number of vortices are large.   After the pioneering work by 
Hess and Fairbank \cite{hes}, and by Packard and Sanders \cite{pac}, the regimes in which only a 
few vortices are present have rarely been explored. Hence, it is not surprising 
that  almost no precise measurement has been made on  $I_z$ just above the $\lambda$ point.

A similar reason exists in theoretical studies as well. 
In theories of superfluidity, the infinite-volume limit is often 
assumed. In $V\rightarrow \infty$,  the only significant distinction 
between states is that between the microscopic and macroscopic one, and therefore 
there is no room for the intermediate-sized wave functions in the theory. 
(In $V\rightarrow \infty$, ``large but not yet macroscopic'' is substantially 
equivalent to ``microscopic'' \cite{equ}.) 
This clear-cut distinction lies behind the two-fluid model, and leads us to 
the preoccupied notion that superfluidity  appear in a mathematically 
discontinuous manner at the Bose-Einstein condensation temperature $T_{BEC}$. 
  For the real system, however, the overall transformation 
 occurs more or less continuously.  For the rotation of 
confined system, one cannot ignore the existence of the center 
of rotation and the boundary of the system, and therefore, one must take 
into account the size of the system.  Hence, the validity of the limit 
 $V\rightarrow \infty$ is worth examination, and the magnitude of phenomena hidden
in the  $V\rightarrow \infty$ limit  must be estimated by 
experiments. 

(1) When one views previous experiments of a liquid helium 4 from this point, one notes in 
the data by Hess and Fairbank an experimental sign suggesting a slight 
decrease of $I_z$ just above $T_{\lambda}$ (see Sec.4A).

(2) The trapped atomic Bose gas opens a new possibility of precise 
measurements of  $I_z$ just above $T_c$. 
The measurement of the angular momentum using the precession of a 
Bose-Einstein condensate of $^{87}$Rb atoms  was performed \cite {che}, which is analogous 
to the experiment by Hess and Fairbank, and by Packard and Sanders in a bulk liquid 
helium 4. It gives a data of $L_z(\Omega)$ like {\it b \/} and {\it c \/} 
in Fig.1 (Fig.2 of Ref.\cite{che}). Since the trapped Bose gas is a 
small system ($\cong 4\mu m$), its $\Omega _c$ is $10^4$ times larger than that 
of a liquid helium 4 in Ref. \cite{hes} and \cite{pac}, which enables us to 
realize a situation in which only a few vortices are present.  Whereas studies in
this field are now centered on the anomalous behaviors at  $T \ll T_c$, it has the 
potential for showing a slight change of $I_z$ just above $T_c$.

Although the deviation of 
the moment of inertia  $I_z$ just above $T_{\lambda}$ from its classical 
value may be small, the essence of superfluidity 
is revealed in a primitive form in such a regime, which constitutes the necessary condition 
for discriminating a quantum fluid from a classical fluid.
To consider these problems, we will make a somewhat different approach from 
conventional ones.  At $T<T_{\lambda}$, the existence of the macroscopic 
coherent wave function $\phi(\mbox{\boldmath $r$})=|\phi|\exp[iS(\mbox{\boldmath 
$r$})]$ ($\mbox{\boldmath $r$}$ is a center-of-mass  
coordinate of many helium 4 atoms) leads to $rot \mbox{\boldmath $v$}=0$ geometrically, because the 
condensate momentum $\mbox{\boldmath $p$}$ is expressed by $\mbox{\boldmath $p$}=(\hbar /m)\nabla  
S$.  Since we will focus on the continuous change of the system around 
$T_{\lambda}$,  we cannot assume from the beginning the sudden emergence of 
$\phi(\mbox{\boldmath $r$})$ at $T_{\lambda}$.  Rather, considering the Bose system above and below $T_{\lambda}$ 
on a common ground will enable us to study the intricacy 
underlying the onset of superfluidity.  We begin with the Bose system 
{\it without the condensate,\/} make a perturbation calculation  of its 
susceptibility with respect to the repulsive interaction by taking 
peculiar graphs reflecting  Bose statistics, and examine  how the formation of the 
coherent wave function gradually changes the rotational behavior of the system. As a result, we derive 
a nonclassical rotational behavior 
that we normally think comes from $rot \mbox{\boldmath $v$}=0$, {\it without 
assuming $rot \mbox{\boldmath $v$}=0$ from the beginning. \/}  
 Specifically, we will derive the decrease of  $I_z$ like {\it d \/} in Fig.1 in the rotating repulsive 
 Bose system just above $T_{\lambda}$, and compare it with the data in a liquid helium 4.

This paper is organized as follows. Section 2 recapitulates the 
definition of the moment of inertia, and explains the physical reason of 
the nonclassical rotational behavior. Section 3 develops a formalism of 
the linear response of the system. (For the difference between this formalism and the 
thermal fluctuation theories, see Appendix.A.)  Using the result in 
Sec.3, Sec.4 re-examines the experiment by Hess and Fairbank, and 
estimates the size of the intermediate-sized wave function and the 
strength of repulsive interaction in a liquid helium 4. 
Section 5 considers the nonlinear response. From this viewpoint, Sec.6 gives a new 
interpretation of the observed decrease of shear viscosity just above 
$T_{\lambda}$ in a liquid helium 4, and discusses some other examples.

\section{Moment of inertia of the repulsive Bose system} 
\subsection{Moment of inertia}  
 Consider bosons in a uniform rotation around   z-axis. For a liquid 
 helium 4, the  repulsive particle picture is not so unrealistic as it would be 
 for any other liquid. Hence, as its simplest model, we use
\begin{equation}
 H=\sum_{p}\epsilon (p)\Phi_{p}^{\dagger}\Phi_{p}
   +U\sum_{p,p'}\sum_{q}\Phi_{p-q}^{\dagger}\Phi_{p'+q}^{\dagger}\Phi_{p'}\Phi_p , 
   (U>0),¥¥¥
	\label{¥}
\end{equation}¥
where $\Phi_{p}$ denotes an annihilation operator of a spinless boson. 

\begin{figure}
\includegraphics [scale=0.3]{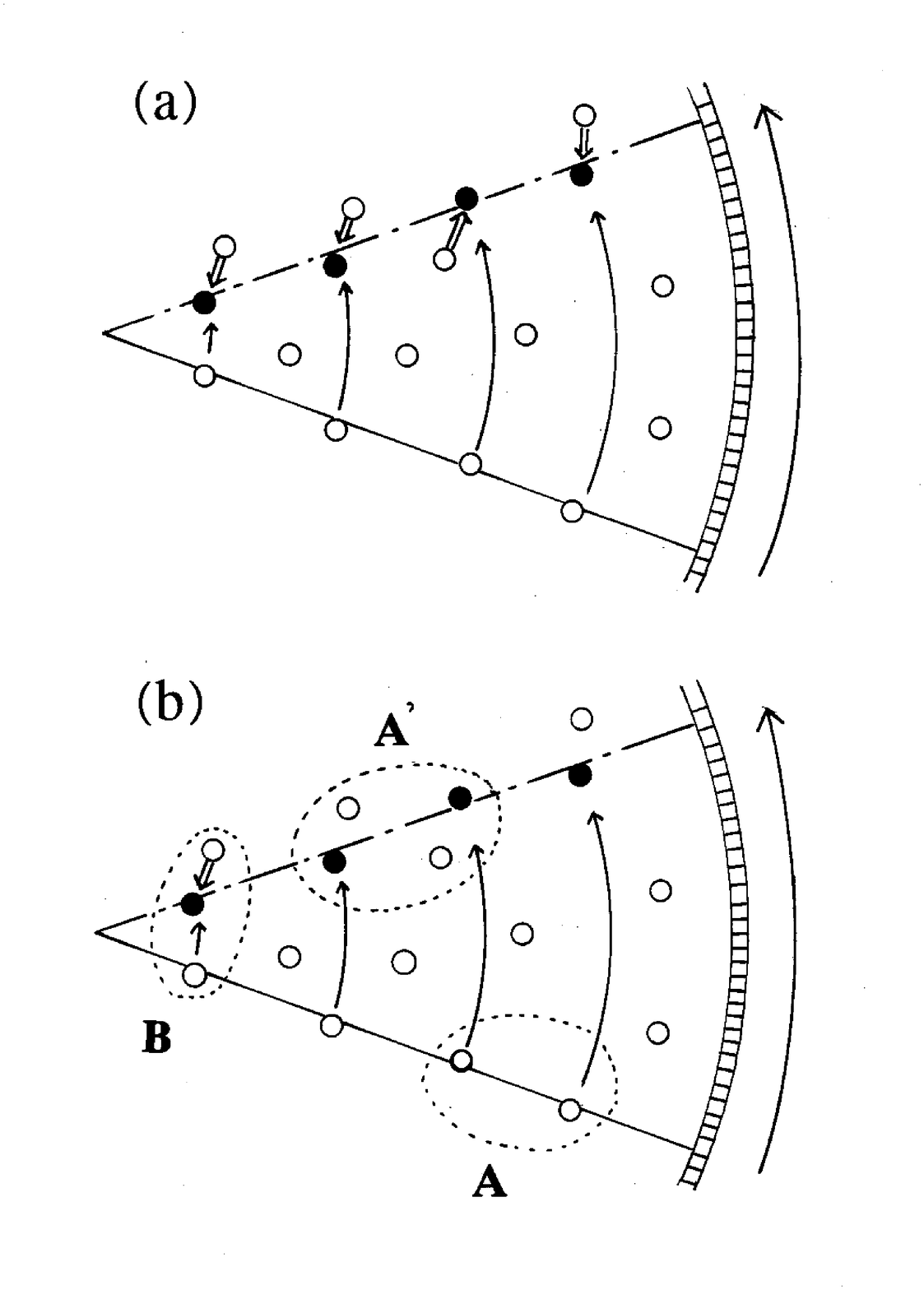}
\caption{\label{fig:epsart}
Schematic pictures of a part of the rotating bosons in a cylindrical container. 
  White circles represent a initial  distribution of particles. The 
  rotation by long arrows moves white circles on a solid-line radius
  to black circles. (a) At $T<T_{\lambda}$, the permutation symmetry holds over the whole liquid. 
 (b) Just above $T_{\lambda}$, it holds only within limited areas 
 enclosed by a dotted lines, a size of which is exaggerated for clarity.}
\end{figure}

 The hamiltonian in a coordinate system  rotating with a container is 
  $H-\mbox{\boldmath $\Omega$}\cdot \mbox{\boldmath $L$}$, where  
$\mbox{\boldmath $L$}$ is the total angular momentum. 
The rotation is equivalent to the application of a probe 
acting on a sample  \cite{noz} \cite{bay}. The perturbation 
$H_{ex}=-\mbox{\boldmath $\Omega$}\cdot \mbox{\boldmath $L$}$ 
is cast in the form $-\sum_{i} (\mbox{\boldmath $\Omega$}\times 
\mbox{\boldmath $r$})\cdot \mbox{\boldmath $p$}¥$ , in which  
$\mbox{\boldmath $\Omega$}\times \mbox{\boldmath $r$} \equiv \mbox{\boldmath $v$}_d( \mbox{\boldmath $r$})$ 
serves as the external field.  Figure.2 shows a part of the Bose system 
in a cylindrical container. When the origin of $\mbox{\boldmath $r$}$ is put on the center 
of rotation,  $\mbox{\boldmath $v$}_d( \mbox{\boldmath $r$})$ has a concentric-circle structure  
illustrated by curved arrows in Fig.2. 
(In the rigid-body rotation, $\mbox{\boldmath $v$}_d( \mbox{\boldmath $r$})$ 
agrees with the drift velocity at point $\mbox{\boldmath $r$}$.) 
We define a mass-current density 
$\mbox{\boldmath $J$}(\mbox{\boldmath $r$})$, and express  the 
perturbation $H_{ex}$ as
\begin{equation}
   -\mbox{\boldmath $\Omega$}\cdot \mbox{\boldmath $L$}
         =-\int  \mbox{\boldmath $v$}_d( \mbox{\boldmath $r$})\cdot 
   \mbox{\boldmath $J$}(\mbox{\boldmath $r$}) d^3x.¥
\end{equation}¥
Because of $ div\mbox{\boldmath $v$}_d( \mbox{\boldmath $r$})=0 $, 
Eq.(2) says that $\mbox{\boldmath $v$}_d( \mbox{\boldmath $r$})$
 acts as a transverse-vector probe to the excitation of bosons. This fact 
 allows us the formal analogy that the response of the system to 
 $\mbox{\boldmath $v$}_d( \mbox{\boldmath $r$})$ is analogous to the 
 response of the charged Bose system to the vector potential 
 $\mbox{\boldmath $A$}(\mbox{\boldmath $r$})$ in the Coulomb gauge.  
 Hence, $ \mbox{\boldmath $J$}(\mbox{\boldmath $r$})$ in Eq.(2) has the 
 following form in momentum space being similar to that in the charged Bose system
 \begin{equation}
	J_{\mu}(q,\tau)=\sum_{p,n} 
	\left(p+\frac{q}{2¥}\right)_{\mu}\Phi_p^{\dagger}\Phi_{p+q}e^{-i\omega _n\tau}¥¥¥,
	\label{¥}
\end{equation}¥
($\hbar =1$ and  $\tau =it$). 
 $\mbox{\boldmath $v$}_d( \mbox{\boldmath $r$})$ is a 
 macroscopic external field  causing the spatial inhomogeneity in the container, 
 whereas $\mbox{\boldmath $J$}(\mbox{\boldmath $r$})$ contains both
 microscopic and macroscopic informations of the system.

 As the simplest susceptibility to $\mbox{\boldmath $v$}_d( \mbox{\boldmath $r$})$, 
  we often use the mass density $\rho =nm$ ($n$ is the number density of particles) as $\mbox{\boldmath $J$} 
  (\mbox{\boldmath $r$})=\rho\mbox{\boldmath $v$}_d( \mbox{\boldmath $r$})$. 
  Microscopically, however, one must begin with the generalized 
  susceptibility consisting of the longitudinal and transverse part ($\mu =x,y,z$)
\begin{equation}
	\chi_{\mu\nu}(q,\omega )=\frac{q_{\mu}q_{\nu}}{q^2¥}\chi^L(q,\omega)     
	            +\left(\delta_{\mu\nu}-\frac{q_{\mu}q_{\nu}}{q^2¥}\right)¥\chi^T(q,\omega) .
	\label{¥}
\end{equation}¥
By definition, the mass density  $\rho$ is a longitudinal response to an external force, $\rho=\chi^L(0,0)$. 
As illustrated in Fig.2, however, the rotational motion of particles is perpendicular 
to the radial direction, along which the influence of the wall motion extends into 
the container. Hence, in principle, one must use the transverse susceptibility $\chi ^T (q,\omega)$ for 
$ \mbox{\boldmath $v$}_d( \mbox{\boldmath $r$} )$ such as 
$\mbox{\boldmath $J$} (\mbox{\boldmath $r$})=\left[\lim_{q\rightarrow 0}
 \chi ^T (q,0)\right] \mbox{\boldmath $v$}_d( \mbox{\boldmath $r$} )$.

Using $ \mbox{\boldmath $\Omega$} =(0,0,\Omega) $ in the left-hand side of Eq.(2) 
and the above $\mbox{\boldmath $J$} (\mbox{\boldmath $r$})$ and 
$\mbox{\boldmath $v$}_d=\mbox{\boldmath $\Omega$}\times \mbox{\boldmath $r$}=(-\Omega y, \Omega x,0) $ 
 in its right-hand side, one obtains the angular momentum $L_z$ as
\begin{equation}
     L_z=  \chi^T(0,0)\int_{V}(x^2+y^2) d^3x \cdot\Omega .¥
\end{equation}¥
In a normal fluid, the susceptibility satisfies $\chi^T(0,0)=\chi 
^L(0,0)$, and therefore the ordinary use of $\rho$ is justified. 
The classical moment of inertia is given by
\begin{equation}
     I_z^{cl}=  mn\int_{V}(x^2+y^2) d^3x = \chi^L(0,0)\int_{V}(x^2+y^2) d^3x .¥
\end{equation}¥
  In a superfluid,  however, the above argument must be altered.
 For the later use, we define a  term proportional to 
 $q_{\mu}q_{\nu}$ in $\chi_{\mu\nu}$ by $\hat{\chi}_{\mu\nu}$
 \begin{eqnarray}
       	\chi_{\mu\nu}(q,\omega)&=&\delta_{\mu\nu}\chi^T(q,\omega)
	                  +q_{\mu}q_{\nu}\left(\frac{\chi^L(q,\omega)-\chi^T(q,\omega)}{q^2¥}\right)¥ \nonumber \\ 
	                          &\equiv& \delta_{\mu\nu}\chi^T(q,\omega)+\hat{\chi}_{\mu\nu}(q,\omega), 
	\label{¥}
\end{eqnarray}¥
where $\hat{\chi}_{\mu\nu}$  represents the balance between the longitudinal and transverse 
susceptibility \cite {tre}. Comparing  Eq.(5) to Eq.(6), one writes the 
moment of inertia $I_z=L_z/\Omega$  using $\hat{\chi}_{\mu\nu}$  
\begin{equation}
      I_z=I_z^{cl}\left(1- \frac{1}{\rho¥}¥\lim_{q\rightarrow 0} 
                      \left[\frac{q^2}{q_{\mu}q_{\nu}¥}\hat{\chi}_{\mu\nu}(q,0)\right]¥\right)¥.
\end{equation}¥
 For the occurrence of nonclassical moment of inertia, the balance between the longitudinal 
and transverse low-energy excitation must be destroyed. In Eq.(8),  $\lim_{q\rightarrow 
0}[(q^2/q_{\mu}q_{\nu})\hat{\chi}_{\mu\nu}(q,0)]$ corresponds to the superfluid 
density $\rho _s$.

 Consider $\chi_{\mu\nu}$ of the ideal Bose system.  Within the linear response, it is defined as
\begin{equation}
     \chi^{(1)}_{\mu\nu}(q,\omega _n)=\frac{1}{V¥}\int_{0}^{\beta¥}d\tau \exp(i\omega_n\tau)
	                      \langle  0|T_{\tau}J_{\mu}(q,\tau)J_{\nu}(q,0)|0\rangle ¥¥¥,
\end{equation}¥
 where $|0>$ is the ground sate of $\sum_{p}\epsilon (p)\Phi_{p}^{\dagger}\Phi_{p}$.  
 The term proportional to $q_{\mu}q_{\nu}$ has a form of
\begin{equation}
\hat{\chi}^{(1)}_{\mu\nu}(q,\omega)  
	            =-\frac{q_{\mu}q_{\nu}}{4¥}¥
	                          \frac{1}{V¥}\sum_{p}\frac{f(\epsilon (p))-f(\epsilon (p+q))}
	                                       {\omega+\epsilon (p)-\epsilon (p+q)¥}¥,
	\label{¥}
\end{equation}¥
where $f(\epsilon (p))$ is the Bose distribution.

(1) If bosons would form the condensate, 
 $f(\epsilon (p))$ in Eq.(10) is a macroscopic 
number for $p=0$ and nearly zero for $p\ne 0$. Thus, in the sum over $p$ in the 
right-hand side of Eq.(10), only two terms corresponding to $p=0$ and 
$p=-q$ remain, with a result that 
\begin{equation}
	\hat{\chi}^{(1)}_{\mu\nu}(q,0)=\rho _s(T)¥\frac{q_{\mu}q_{\nu}}{q^2¥}¥,
\end{equation}¥ 
where $\rho _s(T)=mn_c(T)$ is the thermodynamical superfluid density, and $n_c(T)$ is the 
number density of particles participating in the condensate. 
 Equation.(8) with Eq.(11) leads to
 \begin{equation}
     I_z= I_z^{cl} \left(1-\frac{\rho _s(T)}{\rho¥}¥\right)¥ .¥
\end{equation}¥
(2) When bosons form no condensate, the sum over $p$ in 
Eq.(10) is carried out by replacing it with an integral, and one notes
that $q^{-2}$ dependence disappears in the result. 
Hence, using such a $\hat{\chi}^{(1)}_{\mu\nu}(q,\omega) $ in Eq.(8) 
leads to $I_z=I_z^{cl}$ at $q\rightarrow 0$. This means that, 
without the interaction between particles, BEC is the necessary condition for 
the nonclassical moment of inertia.

Under the repulsive interaction, however, the above argument is seriously 
affected. To see this, we must begin with a physical argument.

\subsection{Bose statistics and repulsive interaction}
We have a physical reason to expect the decrease of the  moment of inertia in 
 bosons at low temperature.  The relationship between 
the low-energy excitations and  Bose statistics dates back to 
Feynman's argument on the scarcity of the excitation in a liquid helium 
4 \cite{fey1}, in which he explained  how Bose 
statistics affects the many-body wave function in configuration space. 
To the rotating bosons, we will apply his explanation.

(1) In Fig.2(a), a liquid (white circles) is in the BEC phase, 
and the wave function has permutation symmetry everywhere in the container.  
Assume that the rotation of a container (depicted by curved arrows) moves 
white circles on a solid-line radius to black circles on a 
one-point-dotted-line radius (a transverse excitation).
At first sight, these displacements seem to be a large-scale configuration change,  
 but this result is reproduced by a set of slight displacements 
(depicted by short thick arrows) from  positions in the initial configuration 
to black circles after rotation.  For any particle after rotation, it is possible 
to find  a particle being close to it in the initial configuration.
 In Bose statistics, owing to permutation symmetry, one cannot 
distinguish between two types of particles after rotation, one  moved 
from the neighboring position by the short arrow, and the other moved 
from distant initial positions by the long arrow. Even if  the 
displacement made by the long arrows is a large 
displacement in classical statistics, it is only a slight 
displacement  by the short arrows in Bose statistics. 

Let us imagine this situation in the 3N-dimensional configuration space. 
The above feature of Bose statistics means that in 
the configuration space, the excited state driven 
by rotation lies close to the ground state. 
Since the excited state is orthogonal to the ground 
state, the wave function corresponding to the excited state must 
spatially oscillate. Accordingly, the many-body wave 
function of the transversely excited state oscillates within a small 
distance in configuration space. Since the kinetic energy of the system 
is determined by the 3N-dimensional gradient of the wave function, this 
steep rise and fall of the amplitude means that the energy of the transverse excitation is 
not small even at $q=0$, leading to the scarcity of the low-energy 
transverse excitation. This is the reason of $\chi^T(q,0)\rightarrow 0$ at $q\rightarrow 0$ 
below $T_{\lambda}$, whereas the particle conservation asserts that 
$\chi^L(q,0)=\rho$ is valid both above and below $T_{\lambda}$. 
Hence, $\hat{\chi}_{\mu\nu}(q,0)$ in Eq.(7) changes to $\rho q_{\mu}q_{\nu}/q^2$ 
at $q\rightarrow 0$, leading to $I_z=0$ in Eq.(8). 
(This mechanism underlies the geometrical condition $rot \mbox{\boldmath $v$}_s=0$.)

(2) At high temperature, the coherent wave function has a microscopic size.  
 If a long arrow  of  $\mbox{\boldmath $v$}_d( \mbox{\boldmath $r$})$ 
 takes a particle to a position beyond the coherent wave function 
 including that particle, one cannot regard the  
 particle after rotation as an equivalent of the initial one.  The mechanism below 
$T_{\lambda}$ does not work for the large displacement extending over two different wave 
functions. Hence, we obtain $\chi^T(q,0)=\chi^L(q,0)$ at $q\rightarrow 0$, and $I_z=I_z^{cl}$.

(3) Figure.2(b) shows the boson system  at the vicinity of $T_{\lambda}$ in the normal 
 phase,  in which the coherent many-body wave function grows to a large but not 
yet a macroscopic size (regions enclosed by a dotted line).  The permutation symmetry holds
only within each of these regions. When  particles are moved from a region A 
to another region A', the mechanism below $T_{\lambda}$ does not work.

 The repulsive interaction $U$ between particles, however, affects this situation. 
 In general, when one moves a particle in the interacting system, 
it induces the motions of other particles. In particular, the
large-distance displacement of a particle in coordinate space causes the 
excitation of many  particles, and therefore it needs a large excitation 
 energy.  This means, in the low-energy excitation of the system, one 
 observes mainly the short-distance displacement of particles.
When applying this tendency to the low-energy excitation of repulsive bosons,
 one knows that {\it excited particles are not likely to go beyond  a single 
coherent wave function, but likely to remain in it, and therefore the mechanism working below 
$T_{\lambda}$ works just above $T_{\lambda}$ as well.\/} This view will be tested as follows. 
If we increase the strength of $U$ in $\chi_{\mu\nu}(q,0)$, the excited 
bosons get to remain in the same coherent wave function, and therefore the low-energy transverse excitation 
 will raises its energy owing  to Bose statistics as discussed in (1). 
 Hence, the condition of $\chi^L(q,0)=\chi^T(q,0) $ at 
$q\rightarrow 0$ will be violated at a certain critical value of $U$. 
Alternatively, {\it if we decrease the temperature at a given $U$, the above condition will be 
violated at a certain temperature $T_{on}$.\/} 

 A geometric feature inherent in the rotation will play an important role in the above mechanism. 
The external field $\mbox{\boldmath $v$}_d(\mbox{\boldmath $r$})$   
 has the structure of concentric circle; hence, the center of rotation is a fixed point. 
 The displacements of particles near the center is so small that they do not go beyond a single  
 coherent wave function (a region B in Fig.2(b)). The 
center of rotation is the most probable point for the mechanism discussed in (a)
 to work.  Hence, the region near the center is most likely to 
decouple from the motion of container. With decreasing temperature, this decoupling will 
 extend from the center to the wall, which depends on the rotational velocity $\Omega$.

To formulate these mechanisms, we consider the perturbation expansion of $\chi_{\mu\nu}$ 
with respect to $U$ in Sec.3. After comparing it with the experiment in 
Sec.4,  we extend it to the nonlinear response to 
$\mbox{\boldmath $v$}_d(\mbox{\boldmath $r$})$ in Sec.5.

\section{Linear response}
 We will formulate the moment of inertia in the repulsive Bose system at 
 the vicinity of $T_{\lambda}$. In the integrand of Eq.(9), one must use, 
 instead of $|0>$, the ground state  $|G>$ of Eq.(1) as follows
 \begin{eqnarray}
	&&\langle G|T_{\tau}J_{\mu}(x,\tau)J_{\nu}(0,0)|G\rangle   \nonumber\\
      &&= \frac{\langle 0|T_{\tau}\hat{J}_{\mu}(x,\tau)\hat{J}_{\nu}(0,0)
 	              exp\left[-\displaystyle {\int_{0}^{\beta¥}}d\tau \hat{H}_I(\tau)¥\right]|0\rangle¥}
 	        {\langle 0|exp\left[-\displaystyle {\int_{0}^{\beta¥}}d\tau  \hat{H}_I(\tau)¥\right]|0\rangle¥},
	\label{¥}
\end{eqnarray}
 where $\hat{H_I}(\tau)$ represents the repulsive interaction. Figure 3(a) 
 illustrates  the current-current response tensor 
 $\hat{J}_{\mu}(x,\tau)\hat{J}_{\nu}(0,0)$ (a lower bubble) in the medium:  
 Owing to $exp(-\smallint \hat{H}_I(\tau)d\tau)$ in Eq.(13), 
 scatterings of particles frequently occur in $|G>$ as illustrated by an upper 
 bubble with a dotted line $U$ in Fig.3(a).  (The black 
and white circle represents a coupling to $\mbox{\boldmath $v$}_d(\mbox{\boldmath $q$})$
 and to $U$, respectively.)  The state $|G>$ includes many interaction 
 bubbles like the upper one in Fig.3(a) with various momentums $p'$ and $p'+q'$.
 A solid line with an arrow represents
\begin{equation}
G(i\omega _n,p)=\frac{1}{i\omega _n-\epsilon(p)-\Sigma+\mu¥}.
	\label{¥}
\end{equation}¥
($\mu$  is a chemical potential implicitly determined by 
$V^{-1}\Sigma[\exp(\beta[\epsilon (q)+\Sigma-\mu])-1]^{-1}=n$).
Owing to the repulsive interaction, the boson has a self energy 
 $\Sigma$ ($>0$) (we ignore its $\omega$ and $p$ dependence by assuming it 
 small). With decreasing temperature, the negative $\mu$ at 
 high temperature approaches a small positive value of $\Sigma$, finally 
 reaching   Bose-Einstein condensation satisfying $\mu=\Sigma$.
 
\begin{figure}
\includegraphics [scale=0.3]{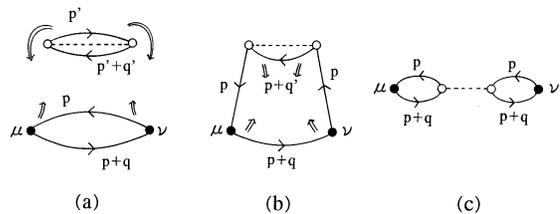}
\caption{\label{fig:epsart}
 (a) The first-order Feynman diagram of a current-current 
  response tensor $J_{\mu}J_{\nu}$ (a lower bubble), and an 
  excitation owing to the repulsive interaction $U$ (an 
  upper bubble with a dotted line).  The black and white small circle  
  represents a vector and scalar vertex, respectively. 
   The exchange of particle lines between the excitation and the  response tensor yields (b). 
    Similarly, the interchange of particle lines in a deformed square  
  yields (c).}
\end{figure}

When the system is just above $T_{\lambda}$ in the normal phase,  
particles in the ground state  $|G>$ are under 
the strong influence of Bose statistics. Hence, the perturbation must be developed 
in such a way that, as the  order of the perturbation increases, the  
susceptibility gradually includes a new effect owing to Bose statistics.
 Specifically, the lower bubble $\hat{J}_{\mu}(x,\tau)\hat{J}_{\nu}(0,0)$ and the upper bubble 
  in Fig.3(a) form a coherent wave function as a whole. 
  When one of the two particles in both the lower and upper bubble have the same momentum 
($p=p'$), and the other in both bubbles have another   
 same momentum ($p+q=p'+q'$) in Fig.3(a), a graph made by 
exchanging these particles must be included in the expansion. Such a
transformation in Fig.3(a) takes place as follows. 
The exchange of two particles having $p$ and $p'$($=p$) by thick white arrows
 yields Fig.3(b). Furthermore, the interchange of two particles having $p+q$ and $p+q'$($=p+q$)
  in Fig.3(b) by thick white arrows  yields 
Fig.3(c). The result is that two bubbles with the same momentum are linked by the 
repulsive interaction, the contribution of which to $\chi_{\mu\nu}$ is given by 
\begin{eqnarray}
&&U\frac{1}{V¥}\sum_{p}(p+\frac{q}{2¥}¥)_{\mu}(p+\frac{q}{2¥}¥)_{\nu}  \nonumber\\
	              &&\times   \left[-\frac{f(\epsilon (p)+\Sigma)-f(\epsilon (p+q)+\Sigma)}
	                            {\omega+\epsilon (p)-\epsilon (p+q)¥} \right]^2¥¥¥.
	\label{¥}
\end{eqnarray}

(a) With decreasing temperature, the coherent wave function grows to a large 
size, and the interchange of particles owing to Bose statistics like Fig.3 occurs 
many times. Hence, one cannot ignore the higher-order terms in Eq.(13), 
which become more significant with the growth of the coherent wave function. 

(b) Among many particles contributing to Eq.(15),  particles stationary to a 
container play a dominant role.  Specifically, a term with $p=0$ in  
 Eq.(15) corresponds to an excitation from the rest 
particle, and  that with $p=-q$ corresponds to a decay into the rest one. 

These two considerations (a) and (b) lead to the following form of 
$\hat{\chi}^{(1)}_{\mu\nu}(q,0)$  at the vicinity of $T_{\lambda}$
\begin{equation}
	\hat{\chi}^{(1)}_{\mu\nu}(q,0)=\frac{q_{\mu}q_{\nu}}{2¥}¥\frac{1}{V¥}\sum_{l=0}^{\infty¥}U^lF_{\beta}(q)^{l+1},
	\label{¥}
\end{equation}¥
where $ F_{\beta}(q)$ is
\begin{equation}
	 \frac{(\exp(\beta[\Sigma-\mu])-1)^{-1}-(\exp(\beta[\epsilon (q)+\Sigma-\mu])-1)^{-1}} 
	                         {\epsilon (q)}¥,¥ 
	\label{¥}
\end{equation}¥
 a positive monotonously decreasing function of $q^2$, which approaches zero as 
$q^2\rightarrow \infty$.  

At a high temperature ($\beta\mu \ll 0$) in which $F_{\beta}(q)$ is 
small,  a small $F_{\beta}(q)$ guarantees the convergence of an infinite series 
in $\hat{\chi}^{(1)}_{\mu\nu}(q,0)$  of Eq.(16), with a result that
\begin{equation}
	\hat{\chi}^{(1)}_{\mu\nu}(q,0)=\frac{q_{\mu}q_{\nu}}{2¥}¥
	                 \frac{1}{V¥}\frac{F_{\beta}(q)}{1-UF_{\beta}(q)¥}.¥¥
	\label{¥}
\end{equation}¥
With decreasing temperature, however, the negative $\mu$ gradually 
approaches $\Sigma$, hence  $\Sigma-\mu\rightarrow 0$. Since  
$F_{\beta}(q)$ increases as $\Sigma-\mu\rightarrow 0$, it   
makes the higher-order term significant in Eq.(16). 
 An expansion form of $F_{\beta}(q)=F_{\beta}(0)-aq^2+\cdots$ 
  around $q^2=0$ has a form such as
 \begin{eqnarray}
 &&F_{\beta}(q) =\frac{\beta}{4\sinh ^2 
	 \displaystyle{\left(\frac{|\beta[\mu(T)-\Sigma]}{2¥}\right)}¥¥¥} \nonumber\\
	                &&\times    \left[1-\frac{\beta}{2¥}\frac{1}{\tanh  \displaystyle{\left(\frac{|\beta[\mu(T)-\Sigma]|}{2¥}¥\right)}¥¥}
	                        \frac{q^2}{2m¥}¥¥¥¥  +\cdots    \right]¥ .¥
	\label{¥}
\end{eqnarray} 
  
  At $q\rightarrow 0$, the denominator $1-UF_{\beta}(q)¥$ in the 
 right-hand side of Eq.(18) 
has a form of $[1-UF_{\beta}(0)]+Uaq^2¥$. In $\Sigma -\mu\rightarrow 0$,
 $UF_{\beta}(0)$ increases and finally reaches 1, that is,  
\begin{equation}
     U\beta=4\sinh ^2\left(\frac{\beta[\mu (T)-\Sigma(U)]}{2¥}¥\right)¥ .
	\label{¥}
\end{equation}¥
At this point, the denominator in the right-hand side of Eq.(18) gets to begin with $q^2$, and 
$\hat{\chi}_{\mu\nu}^{(1)}(q,0)$ therefore changes to a form of 
$q_{\mu}q_{\nu}/q^2$  at $q\rightarrow 0$.
 This means that a non-zero coefficient $F_{\beta}(0)/(2VUa)$ of 
 $q_{\mu}q_{\nu}/q^2$ appearing in Eq.(18) gives
  a non-zero value of $\chi^L(q,0)-\chi^T(q,0)$ in Eq.(7), hence the 
  moment of inertia shows  the 
nonclassical behavior in Eq.(8).

 From now, we call $T$ satisfying Eq.(20)  {\it the onset temperature of the 
 nonclassical moment of inertia \/}  $T_{on}$ \cite {com}. 
 
 At the vicinity of $T_{\lambda}$, Eq.(20) is approximated as $ 
 U\beta=\beta^2[\mu (T)-\Sigma(U)]^2$ for a small $\mu-\Sigma$. This 
 condition has two solutions $\mu (T)=\Sigma(U)\pm \sqrt{Uk_BT}$. It is 
 generally assumed that the repulsive Bose system undergoes BEC 
  as well as a free Bose gas. Hence, with decreasing temperature, $\mu 
  (T)$ in the presence of repulsive interaction $U$
 should reach $\Sigma (U)$ at a finite temperature, during which course the system necessarily passes a 
 state satisfying  $\mu (T)=\Sigma (U)-\sqrt{Uk_BT}¥$ \cite{att}. One 
 concludes that {\it the nonclassical rotational behavior always occurs prior 
 to BEC in the repulsive bosons \/}, that is, $T_{on}>T_{\lambda}$.
 
 The chemical potential $\mu$, hence $\mu -\Sigma$ as well, determines 
 the size of the coherent many-body wave function \cite{mat}\cite{fey2}, which 
 corresponds to the size of regions enclosed by a dotted line in Fig.2(b). {\it The 
 emergence of $q^{-2}$ singularity in  $\hat{\chi}_{\mu\nu}^{(1)}(q,0)$ 
 in the process of $\Sigma -\mu\rightarrow 0$ is a mathematical expression of the 
 instability mechanism induced by the growth of the coherent wave function. \/}
 When $U$ is small, this instability occurs after the wave function 
 grows to a large size corresponding to a small $\Sigma -\mu$. When $U$ is 
 large, this instability already occurs at a larger $\Sigma -\mu$ in which the wave 
 function is smaller than the former one.
 
  At the onset temperature  $T_{on}$, substituting Eq.(19) into 
  Eq.(18), we find $\hat{\chi}^{(1)}_{\mu\nu}$ at $q\rightarrow 0$
 \begin{equation}
	\hat{\chi}^{(1)}_{\mu\nu}(q,0)=\frac{2m}{U\beta_{on}¥}\frac{1}{V¥}\tanh \left(\frac{|\beta _{on}[\mu(T_{on})-\Sigma]|}{2¥}¥\right)¥¥
	                              \frac{q_{\mu}q_{\nu}}{q^2¥}¥ ,
\end{equation}¥
and  with the aid of Eq.(20)
\begin{equation}
	\hat{\chi}^{(1)}_{\mu\nu}(q,0)=\frac{1}{V¥}\frac{m}{\sinh |\beta _{on}[\mu(T_{on})-\Sigma] |¥}¥
	                              \frac{q_{\mu}q_{\nu}}{q^2¥}¥.
\end{equation}¥
$\hat{\chi}^{(1)}_{\mu\nu}(q,0)$ is given by
\begin{equation}
	\hat{\chi}^{(1)}_{\mu\nu}(q,0)=mc(T_{on})n_0(T_{on})\frac{q_{\mu}q_{\nu}}{q^2¥},
\end{equation}¥
where 
\begin{equation}
     n_0(T)=\frac{1}{V¥}\frac{1}{\exp(-\beta [\mu(T)-\Sigma])-1¥}¥¥,
	\label{¥}
\end{equation}¥
is the number density of $p=0$ bosons, and
\begin{equation}
	c(T)=\frac{2}{\exp(\beta |\mu(T)-\Sigma|)+1¥}¥,
\end{equation}¥
 is a  Fermi-distribution-like coefficient. For the finite system just above $T_{\lambda}$, 
 $n_0(T)$ has a large but not yet macroscopic value.
In the theoretical limit $V\rightarrow\infty $, this quantity is normally 
regarded to be zero. In real finite system, however,  its magnitude must 
be estimated by experiments (see Sec.4).  Using Eq.(23) in Eq.(8), we obtain 
\begin{equation}
     I_z(T_{on})=I_z^{cl} \left(1-c(T_{on})\frac{n_0(T_{on})}{n¥}¥\right)\equiv I_z^{cl} \left(1-\frac{\hat{\rho _s}(T_{on})}{\rho¥}¥\right)¥ .¥
\end{equation}¥
where $\hat{\rho _s}(T) \equiv mc(T)n_0(T)$ is  {\it the mechanical superfluid density\/}. 

$\hat{\rho _s}(T)$ reflects the intermediate-sized coherent wave function. 
``Intermediate'' means that it does not play the role of order parameter characterizing the 
thermodynamical phase, but affects mechanical properties of the system.
  At $T=T_{on}$, the system shows a small but finite jump from 
$I_z^{cl}$ to $I_z$, which is proportional to $n_0(T_{on})$.
Since $c(T_{on})$ has an order of one, the magnitude of this jump  
is determined  mainly by the non-macroscopic $n_0(T_{on})$ and 
 the moment of inertia therefore only slightly decreases from 
 the classical value. (This does not mean that the thermodynamical 
 quantities show a finite jump.)  
At $T<T_{on}$, the moment of inertia varies with $T$ following $I_z(T)$ 
in Eq.(26). Equation (26) determines an initial slope of the dotted curve 
{\it d \/} at $\Omega =0$ in Fig.1.
 When the  system reaches $T=T_{\lambda}$, the condition of  $\mu=\Sigma$ makes 
 $\hat{\rho _s}(T_{\lambda})=\rho _s(T_{\lambda})$ 
 because of $c(T_{\lambda})=1$, $n_0(T_{\lambda})=n_c$, 
which shows a natural connection of Eq.(26) to Eq.(12) with the 
thermodynamical $\rho _s(T)$. While the thermodynamical $\rho _s(T)$
  satisfy $\rho _s=(m^2k_BT/\hbar ^2) |\phi|^2$ at $T<T_{\lambda}$, one can expect 
no simple relation of the mechanical $\hat{\rho _s}(T)$ with the 
thermodynamical quantities at $T_{\lambda}<T<T_{on}$.

\section{Comparison to experiments}
 \subsection{Experiment by Hess and Fairbank revisited}
Hess and Fairbank, after they confirmed that a superfluid 4 below 
$T_{\lambda}$ remained at rest under the extremely slow rotation, made 
another type of experiment in their classic paper Ref.\cite{hes}.  
First,  at an initial temperature $T_1$ below or above  $T_{\lambda}$, 
they rotated a liquid helium 4, contained in a small cylinder of radius 
$R=$ 0.44 mm, at $\Omega$ =1.13 rad/s. 
Later they  heated it up to the temperature as it comes into rigid-body rotation, 
and precisely measured a small change  $\Delta\Omega$ of the angular velocity. 
 Using Eq.(12), the rotational energy before heating is $1/2\times 
 I_z^{cl}(1-\rho _s/\rho)\Omega^2$, 
 whereas it changes to $1/2\times I_z^{cl}(\Omega -\Delta\Omega)^2$ after heating. 
 By the conservation of energy, the former must be equal to the latter, 
with a result that  $2\Delta\Omega /\Omega\cong \rho _s/\rho$.
 Below $T_{\lambda}$, a fraction of liquid does not participate in 
the rotation, whereas after heating it does.  The 
rotation after heating therefore always becomes slower, and 
$2\Delta\Omega /\Omega\cong  \rho _s/\rho$ is 
positive. They plotted  $\Delta\Omega /\Omega$ as a function of initial 
 temperature  $T_1$ (Fig.2 of Ref.\cite{hes}).

When the initial temperature $T_1$ was lower than $T_{\lambda}$,  
$\Delta\Omega /\Omega$ was properly explained by the 
theoretical value of $\rho _s(T_1)/\rho$.  If the two-fluid model is 
exactly valid near $T_{\lambda}$, $\rho _s$ must vanish at 
$T>T_{\lambda}$. Hence, $\Delta\Omega /\Omega$ measured under the 
condition of $T_1>T_{\lambda}$ must be exactly zero. 
 For $T_1=T_{\lambda}+0.03K$ and $T_{\lambda}+0.28K$, however, the measured $\Delta\Omega /\Omega$ in 
Ref \cite {hes} were not exactly zero. Although the error bars were large 
compared with its absolute values, 
its central values were significantly different from zero:  $\Delta\Omega 
/\Omega\cong 4\times 10^{-5}$ at $T_1=T_{\lambda}+0.03K$, and  $1.5\times 
10^{-5}$ at $T_{\lambda}+0.28K$. 
A natural interpretation of this result is that these $T_1$'s 
were lower than the onset temperature $T_{on}$ of the nonclassical moment of 
inertia, and at such $T_1$'s, $I_z$ was already slightly smaller than its classical value.
If it is true, instead of Eq.(12), $I_z(T)=I_z^{cl}(1-\hat{\rho _s}(T)/ \rho)$ 
(Eq.(26)) must be used, and we obtain  
 $2\Delta\Omega /\Omega\cong \hat{\rho _s}(T_1)/\rho$. Hence, we obtain 
 $\hat{\rho _s}(T_{\lambda}+0.03K)/\rho\cong 8\times 10^{-5}$, and  
 $\hat{\rho _s}(T_{\lambda}+0.28K)/\rho\cong 3\times 10^{-5}$ \cite{den}. 
 Since the number density of atoms in a liquid helium 4 is 
 $n\cong 2.2\times 10^{22}$ atoms/cm$^{3}$, we have $n_0\cong 10^{18}$ atoms/cm$^{3}$. 
  This is an experimental estimation of the number of helium 4 atoms participating 
  in the intermediate-sized coherent wave function in a bulk helium 4 just above  
  $T_{\lambda}$. Two temperatures,
$T_{\lambda}$+0.03K and $T_{\lambda}$+0.28K, are situated within the temperature region  
($T_{\lambda}<T<$ 2.8 K) in which the viscosity begins to decrease above  $T_{\lambda}$. 
In Ref.\cite{hes}, the author's focus was on the rigidity 
of a superfluid against the rotation (the result was later named Hess-Fairbank effect), 
and they did not mention the small non-zero value of $\Delta\Omega /\Omega$ above  
$T_{\lambda}$. Although this point did not attract the interest of many 
people, it is worth studying closely in the future.

\subsection{Estimation of the repulsive interaction $U$}
 Let us make an estimation of the repulsive interaction $U$  using the 
 approximate form of Eq.(20), $U\beta _{on}=\beta _{on} ^2(\mu -\Sigma)^2$. 
 (1) For the present, we suppose $T_{on}=2.8$K by assuming that the 
 decrease of viscosity just above $T_{\lambda}$   
has a relation to the emergence of the intermediate-sized 
 coherent wave function (see Sec.6.A). (2) We assume  that  $\mu(T) -\Sigma (U)$ in 
 Eq.(20) follows the formula
\begin{equation}
	\mu (T)-\Sigma (U)=-\left(\frac{g_{3/2}(1)}{2\sqrt{\pi}¥}¥\right)^2k_BT_{\lambda}
	          \left[\left(\frac{T}{T_{\lambda}¥}\right)^{3/2}-1\right]^2,
	\label{¥}
\end{equation}¥ 
($g_{a}(x)=\sum_{n}x^n/n^{a}¥$) on the assumption that the particle 
interaction $U$ and the particle density of a liquid helium 4 
  are renormalized to $T_{\lambda}=2.17K$ (an approximation that dates back to London). 
  Hence, we obtain a rough estimation of $U$ as $0.5 \times 
 10^{-17}$ erg. This value is approximately close to the repulsive 
 interaction $U_c$ obtained by Bogoliubov's  spectrum $c=(\hbar 
 /m)\sqrt{4\pi na}$, but somewhat smaller than it. 
  (The velocity of ordinary sound $c=220$ 
 $m/s$ of a liquid helium 4 near $T_{\lambda}$ gives a scattering length $a=0.7$ 
 $nm$, hence  $U_c\cong \hbar ^2/(ma^2)=3.4\times 10^{-17}$ erg.) It is a 
 difficult problem to relate these values to the realistic potential 
 between helium 4 atoms such as the Lennard-Jones potential
  $U(r)=4\epsilon[(2.556/r)^{12}-(2.556/r)^{6}]$ with $\epsilon =1.41 \times 10^{-15}$ erg.

 \section{Nonlinear response}
 When the precise measurement of $I_z$ is performed, the dynamic response of 
 $I_z(\Omega)$  will become a next subject, which  appears in the 
 nonlinear response of the system.  
   As discussed in Sec.2,  
 $\mbox{\boldmath $J$}$ contains both microscopic and macroscopic information, whereas 
 $\mbox{\boldmath $v$}_d( \mbox{\boldmath $r$})$ is a macroscopic external field, 
 and therefore the susceptibility connecting $\mbox{\boldmath $J$}$ and 
 $\mbox{\boldmath $v$}_d$ appears  
 as $\mbox{\boldmath $J$} (\mbox{\boldmath $r$})=\left[\lim_{q\rightarrow 0}
 \chi ^T (q,0)\right] \mbox{\boldmath $v$}_d( \mbox{\boldmath $r$} )$. 
Hence, we  consider not a general form of the nonlinear 
susceptibility, but  a correction term $\chi^{T,non}(v_d)$ to the linear 
response, such as $\mbox{\boldmath $J$}=[\chi^{T,(1)}+\chi^{T,non}(v_d)]\mbox{\boldmath $v$}_d$.
 Since $\chi^{T,non}(v_d)$ does not depend on the direction of $\mbox{\boldmath $v$}_d$, 
it includes only even powers of $v_d=\Omega r$, which 
leads to the spatial inhomogeneity and the dynamic response.

For the dynamic response, we define some 
 quantities. The current $\mbox{\boldmath $J$}(\mbox{\boldmath $r$})=\chi^T(0,0)
\mbox{\boldmath $v$}_d( \mbox{\boldmath $r$})$ in Sec.2.A is replaced by
 $\mbox{\boldmath $J$}(\mbox{\boldmath $r$})=\chi^T(0,0,\Omega, r)
\mbox{\boldmath $v$}_d( \mbox{\boldmath $r$})$, where 
 $r$=$\sqrt{x^2+y^2}¥$ is a distance from the center of rotation. 
Correspondingly, instead of Eq.(5) and (8), we define
\begin{equation}
     L_z= \int_{V}\chi^T(0,0, \Omega, r)r^2 d^3x \cdot \Omega ,¥
\end{equation}¥
and
\begin{equation}
     I_z(\Omega)= I_z^{cl} - \lim_{q\rightarrow 0} \int_{V}
                      \left[\frac{q^2}{q_{\mu}q_{\nu}¥}\hat{\chi}_{\mu\nu}(q,0,\Omega, r)\right]¥
                         r^2 d^3x .¥
\end{equation}¥
The position-dependent rotational velocity is defined as
\begin{equation}
     \mbox{\boldmath $\Omega$}_0(\mbox{\boldmath $r$})= 
                   \left(1-\frac{1}{\rho¥}¥\lim_{q\rightarrow 0} 
                   \left[\frac{q^2}{q_{\mu}q_{\nu}¥}\hat{\chi}_{\mu\nu}(q,0,\Omega,  
                   r)\right]\right)¥¥\mbox{\boldmath $\Omega$}.
\end{equation}¥
We extend the mechanical superfluid density $\hat{\rho _s}(T)$ in Eq.(26) so 
that it has $\Omega$ and $r$ dependence  and  satisfies
$\hat{\rho _s}(T,\Omega)=\int \hat{\rho _s}(T,\Omega,r)2\pi rdrdz¥$ as
\begin{equation}
       \hat{\rho _s}(T,\Omega,r)
               =\lim_{q\rightarrow 0} \left[\frac{q^2}{q_{\mu}q_{\nu}¥}\hat{\chi}_{\mu\nu}(q,0,\Omega, r)\right]  .¥
\end{equation}¥

   Let us consider the first approximation of the above quantities. 
  We begin  with 
   \begin{equation}
   	 <J_{\mu}(x,t)>= <G|S^{\dagger}\hat{J}_{\mu}(x,t)S|G>,
   	\label{¥}
   \end{equation}¥
where $S= T exp\left[-i\int_{-\infty}^{t}dt' \hat{H}_{ex}(\mbox{\boldmath $r$},t')¥\right]¥$.
  Using $H_{ex}(\mbox{\boldmath $r$})=-v^{\mu}_{d}(\mbox{\boldmath 
  $r$})J_{\mu}(\mbox{\boldmath $r$})$, the analytical continuation 
  $t\rightarrow \tau =it$ is performed in the higher-order expansion terms 
 in the right-hand side of Eq.(32) \cite{sle}. 
 As the simplest nonlinear susceptibility for $J_{\mu}$, we consider the third-order term 
 $\chi^{(3)}_{\mu,\nu\sigma\tau}v_d^{\nu}v_d^{\sigma}v_d^{\tau}$ with 
 respect to $H_{ex}$ and extract  a correction term to the linear  
 susceptibility  $\chi^{(3)}_{\mu\nu}(v_d)\mbox{\boldmath $v$}_d$ from  
 $\chi^{(3)}_{\mu,\nu\sigma\tau}v_d^{\nu}v_d^{\sigma}v_d^{\tau}$.
$\chi^{(3)}_{\mu\nu}(v_d)$ is illustrated as a square in Fig.4(a), in which 
 we choose only one vertex out of three vertices for the coupling to $\mbox{\boldmath $v$}_d( 
 \mbox{\boldmath $q$})$ and make the others couple to the coarse-grained external field 
 $\mbox{\boldmath $v$}_d( \mbox{\boldmath $r$})$. The lowest-order dynamic 
response of $I_z(\Omega)$ is obtained through $\chi^{(3)}_{\mu\nu}(v_d)$. (From 
now, we express $\chi^{(3)}_{\mu\nu}(v_d)$ by $\chi^{(3)}_{\mu\nu}(q,i\omega)$.)

\begin{figure}
\includegraphics [scale=0.3]{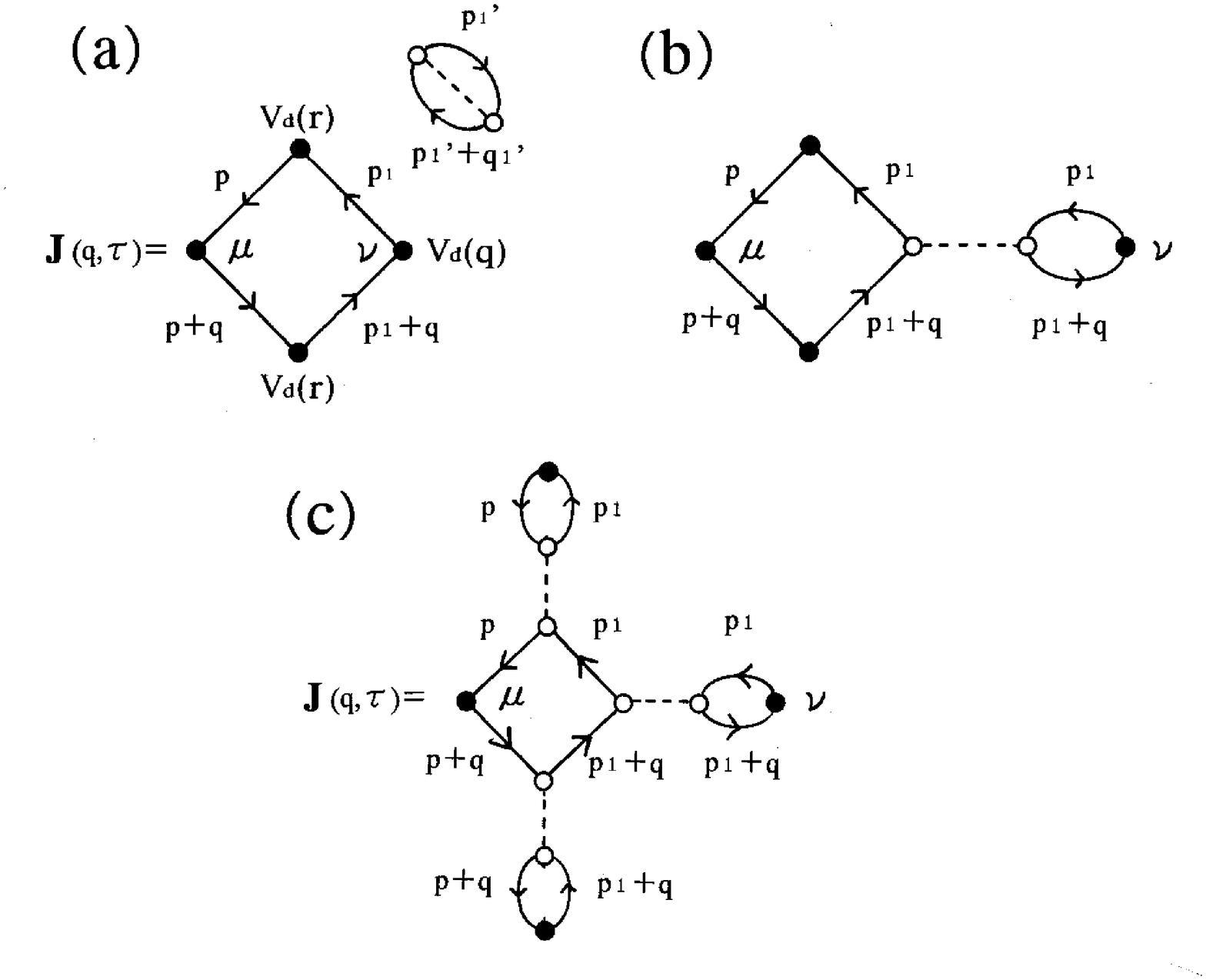}
\caption{\label{fig:epsart}
 (a) The third-order Feynman diagram of the current-current 
   response tensor accompanied by the bubble excitation. (The 
   definition of symbols is same as in Fig.3.) The exchange 
   of particle at one vertex yields (b). Similar exchanges 
  at the other vertices yield (c). }
\end{figure}

Comparing Fig.3(a) and 4(a), we find that $p$ in the 
lower bubble of Fig.3(a) splits into $p$ and $p_1$ in Fig.4(a). 
We obtain a formula corresponding to the square in Fig.4(a)
\begin{eqnarray}
     \chi^{(3)}_{\mu\nu}(q,i\omega)
     &=& \beta n_0|\mbox{\boldmath $v$}_d( \mbox{\boldmath $r$})|^2¥
       \frac{1}{\beta  ^2¥¥}¥\sum_{n,m}¥\frac{1}{V^2¥}\sum_{p,p_1}\left(p+\frac{q}{2¥}¥\right)_{\mu}   \nonumber\\ 
     & \times &\left(p_1+\frac{q}{2¥}¥\right)_{\nu}
                  \left(\frac{\mbox{\boldmath $p$}+\mbox{\boldmath $p_1$}}{2¥}¥\right)\cdot 
                  \left(\frac{\mbox{\boldmath $p$}+\mbox{\boldmath $p_1$}}{2¥}+\mbox{\boldmath $q$}¥\right)        \nonumber\\ 
     & \times & G(i\omega _n+i\omega,p+q) G(i\omega  _n,p)                                  \nonumber\\ 
     & \times & G(i\omega _m+i\omega,p_1+q) G(i\omega _m,p_1),
	\label{¥}
\end{eqnarray}¥
where $[(\mbox{\boldmath $p$}+\mbox{\boldmath $p_1$})/2]\cdot 
[(\mbox{\boldmath $p$}+\mbox{\boldmath $p_1$})/2+\mbox{\boldmath $q$}]$ comes from the coupling to the upper 
and lower $\mbox{\boldmath $v$}_d( \mbox{\boldmath $r$})$ in Fig.4(a). 
In general, a loop with four vertices has three inner frequencies 
$\omega_n, \omega_m$, and $\omega_l$. For the susceptibility like Fig.4(a), 
however, the $q,\omega$ in $\chi_{\mu\nu}(q,i\omega)$ enters at one of the four vertices and leaves 
at another, thus leaving only two frequencies $\omega_n$, $\omega_m$ as 
internal ones in Eq.(33).  
The remaining $\hat{H}_{ex}$ in $\int_{0}^{\beta¥}¥ d\tau \hat{H}_{ex}$, 
corresponding to $\omega _l$, appears as 
$\beta\hat{H}_{ex}$ ($=\beta n_0|\mbox{\boldmath $v$}_d( \mbox{\boldmath $r$})|^2$), 
since the macroscopic $\hat{v}_d(r)$ in $\hat{H}_{ex}$  slowly varies with $\tau$.

(a) With decreasing temperature, the coherent wave functions gradually grow, 
and therefore the particle interchange owing to Bose statistics frequently occurs in 
$\chi^{(3)}_{\mu\nu}$.  By applying Eq.(13)-like formula for the product 
of four currents,  we obtain a perturbation expansion of 
$\chi^{(3)}_{\mu\nu}$ with respect to the repulsive 
interaction  $H_{I}$. Figure.4(a) illustrates the square 
 surrounded by a bubble in $|G>$ (an analogue of Fig.3(a)).  
 The  particle interchange  at one of three  vertices yields Fig.4(b). 
 Fig.4(c) shows a horizontal and vertical extension of bubble chains. 

(b)  With decreasing temperature, particles stationary to a container 
get to play a dominant role. Comparing Fig.3(c) and Fig.4(b), 
we apply the argument above Eq.(15) to Eq.(33), knowing that $p=0$ and 
$p=-q$ plays a dominant role in the sum over $p$.  In the sum over $p_1$ as well, 
the dominant process comes from $p_1=0$ and $p_1=-q$.
At first sight, there seems to be four combinations in $(p,p_1)$.
Owing to $(\mbox{\boldmath $p$}+\mbox{\boldmath $p_1$})/2$ and 
$(\mbox{\boldmath $p$}+\mbox{\boldmath $p_1$})/2+\mbox{\boldmath $q$}$ in Eq.(33),
 however,  only $(p,p_1)=(0,-q)$ and $(-q,0)$ are possible. 
In Fig.4(a), the coupling to $v_d(q)$ is possible on the upper or lower vertex 
 as well. Although this case has a different expression of 
$\chi^{(3)}_{\mu\nu}(q,i\omega)$ from Eq.(33), the particle 
interchange and the dominance of $p=0$ or $p=-q$ particles 
derive the same formula from its $\chi^{(3)}_{\mu\nu}(q,i\omega)$, 
giving the final result the symmetry factor 2. 
Hence, one obtains the term representing the balance between the longitudinal and transverse 
susceptibility for the square in Fig.4(a)
\begin{equation}
	\hat{\chi}^{(3)}_{\mu\nu}(q,0,\Omega,r)= -q_{\mu}q_{\nu}\frac{1}{V^2¥}
	               \left(\frac{q^2}{4}\right)¥ 
	               |F_{\beta}(q)|^2 (\Omega r)^2\beta n_0(T). ¥
	\label{¥}
\end{equation}¥

The process from  Fig.4(a) to (c) changes Eq.(34) as follows.  Using Eq.(13)-like 
formula for the product of four currents, bubble 
 chains in Fig.4(c) are extended to infinity. Among these terms, to derive the nonclassical 
 $I_z(\Omega)$ in Eq.(29), we pick up only terms that give 
 a factor proportional to $q_{\mu}q_{\nu}/q^2$ at $q\rightarrow 0$ limit  
 because other terms vanish in the $q\rightarrow 0$ limit of Eq.(29). Hence, 
 \begin{eqnarray}
	&&\hat{\chi}^{(3)}_{\mu\nu}(q,0,\Omega,r)= -\frac{q_{\mu}q_{\nu}}{2}\left(\frac{q^2}{2}\right)\frac{1}{V^2¥}\frac{1}{n^2¥}  \nonumber\\
	    && \times \frac{|F_{\beta}(q)|^2}{\left[1-UF_{\beta}(q)\right]^2¥}(\Omega r)^2\beta  n_0(T)¥. ¥
	\label{¥}
\end{eqnarray} 

Similarly to Eq.(18), the instability occurs in Eq.(35)  
when the condition of Eq.(20) is satisfied, because the denominator in the right-hand 
side of Eq.(35) gives $q^{-4}$, and therefore the coefficient of 
$q_{\mu}q_{\nu}$ diverges as $q^{-2}$ at $q=0$.   At $T=T_{on}$, the same 
procedure as that from Eq.(18) to (21) yields
\begin{eqnarray}
  &&\hat{\chi}^{(3)}_{\mu\nu}(q,0,\Omega,r)= -\frac{1}{V^2¥}\left(\frac{2m}{U\beta_{on}}\right)^2 \beta_{on}\frac{1}{n^2¥}n_0(T_{on})¥ \nonumber\\
               	&&\times 	\tanh ^2 \left(\frac{\beta_{on}[\mu(T_{on})-\Sigma]}{2¥}¥\right)¥ 
                (\Omega r)^2  \frac{q_{\mu}q_{\nu}}{q^2}¥,
	\label{¥}	
\end{eqnarray} 
and  with the aid of Eq.(20)
\begin{eqnarray}
	&&\hat{\chi}^{(3)}_{\mu\nu}(q,0,\Omega, r) = -\frac{mn_0(T_{on})}{V^2¥}\left(\frac{m(\Omega r)^2}{k_BT_{on}}\right)¥ \nonumber\\
                &&\times  \frac{1}{n^2¥}\frac{1}{\sinh ^2(\beta_{on}[\mu(T_{on})-\Sigma])¥} \frac{q_{\mu}q_{\nu}}{q^2}¥.
	\label{¥}		
\end{eqnarray} 
 Using $\hat{\rho _s}(T)= mc(T)n_0(T)$ as in Eq.(23), we obtain
\begin{equation}
	\hat{\chi}^{(3)}_{\mu\nu}(q,0,\Omega, r)= -mn_0(T_{on})\left(\frac{\hat{\rho _s}(T_{on})}{\rho}\right)^2
	                              \left(\frac{m(\Omega r)^2}{k_BT_{on}}\right)¥ \frac{q_{\mu}q_{\nu}}{q^2}¥.
	\label{¥}
\end{equation}¥

Substituting Eqs.(23) and (38) to Eq.(30) and using $mn_0/\rho =c^{-1}(\hat{\rho 
_s}/\rho)$ from the definition of $\hat{\rho _s}$, we get at $T \le T_{on}$
\begin{eqnarray}
	 && \mbox{\boldmath $\Omega$}_0(\mbox{\boldmath $r$}) \nonumber\\
       &&=  \left[1-\frac{\hat{\rho _s}(T)}{\rho¥}¥+\frac{1}{c(T)¥}¥¥\left(\frac{\hat{\rho _s}(T)}{\rho¥}¥\right)^3
                         \frac{m(\Omega r)^2}{k_BT¥}¥\right]¥
                           ¥¥\mbox{\boldmath $\Omega$}.	
\end{eqnarray} 
 Using Eq.(39), we obtain a velocity field $\mbox{\boldmath $v$}_0( \mbox{\boldmath $r$})=
\mbox{\boldmath $\Omega$}_0(\mbox{\boldmath $r$})\times \mbox{\boldmath 
$r$}$, and a vorticity field
\begin{equation}
    rot \mbox{\boldmath $v$}_0 (\mbox{\boldmath $r$}) = 2\Omega
             \left[1-\frac{\hat{\rho _s}(T)}{\rho¥}¥+\frac{2}{c(T)¥}\left(\frac{\hat{\rho _s}(T)}{\rho¥}¥\right)^3
                         \frac{m(\Omega r)^2}{k_BT¥}¥\right]
             \mbox{\boldmath $e$}_z.¥
\end{equation}¥
This form shows a little change from $ rot \mbox{\boldmath $v$}=2\Omega\mbox{\boldmath $e$}_z$ 
to $rot \mbox{\boldmath $v$}=0$, that is, from a normal fluid to a 
superfluid. These results are summarized in the mechanical superfluid density
\begin{equation}
    \hat{\rho _s}(T,\Omega,r)=\hat{\rho _s}(T) \left[1-\frac{1}{c(T)¥}\left(\frac{\hat{\rho _s}(T)}{\rho¥}¥\right)^2
                         \frac{m(\Omega r)^2}{k_BT¥}\right]¥ .
\end{equation}¥
In a liquid helium 4, $m/k_{B} \cong 2.13\times 10^{-8}$ sec$^2/$cm$^2$. Hence, 
 the $\Omega$ dependence of Eq.(41) is negligibly small.

 When the container is rotated in the normal phase, a liquid makes the rigid-body rotation 
satisfying $\mbox{\boldmath $v$}_d( \mbox{\boldmath $r$})=
\mbox{\boldmath $\Omega$}\times \mbox{\boldmath $r$}$.  In the superfluid phase, a quantum 
vortex appears and a liquid makes a peculiar rotation 
 satisfying $\mbox{\boldmath $v$}( \mbox{\boldmath $r$})=(\hbar l/m) \mbox{\boldmath 
$e$}_z\times  \mbox{\boldmath $r$}/r^2$ ($l$ is an integer) being independent of $\Omega$. At 
$T_{\lambda}<T<T_{on}$, however, while a liquid follows the 
rotation of a container, a region around the center of 
rotation slightly reduces its rotational velocity. Hence, the velocity 
field in a container takes an intermediate form between the velocity near the center and at the boundary. 
One can call it a {\it differential rotation \/} \cite {dif}. 

(1) Near the boundary,  $\mbox{\boldmath $v$}_d( \mbox{\boldmath $r$})=
\mbox{\boldmath $\Omega$}\times \mbox{\boldmath $r$}$ becomes a 
 large external field, and more higher-order terms than $\chi^{(3)}$ will contribute to 
 $I_z(\Omega)$.  Presumably, this will suppress superfluidity, and $\mbox{\boldmath 
 $v$}_0( \mbox{\boldmath $r$})=\mbox{\boldmath $\Omega$}_0( \mbox{\boldmath $r$})\times 
 \mbox{\boldmath $r$}$ with Eq.(39) will approach $\mbox{\boldmath $v$}_d( \mbox{\boldmath $r$})=
\mbox{\boldmath $\Omega$}\times \mbox{\boldmath $r$}$ of 
 the rigid-body rotation.  
 
 (2) When one increases the rotational velocity $\Omega$ 
 of a container, one can expect a similar result. 
   With increasing $\Omega$, the region rotating more slowly than the container
 will shrink to the center of rotation, and therefore 
 the nonclassical $I_z$ will approach the classical one  as illustrated 
 by a dotted line {\it d \/}in Fig.1. 
 
 (3) As $T\rightarrow T_{\lambda}$, the region near the center of rotation, which has a smaller vorticity,  
 will enlarge toward the wall. (To describe such a growth and shrink, the effect of the 
wall must be taken into account as a boundary condition.) 

 (4) At $T<T_{\lambda}$,  the nonclassical behavior discussed in this paper 
is masked by the quantum jump of $L_z$  owing to the emergence of the quantum vortex. 
 If the quantum vortex could be suppressed, one would see the hidden 
 behavior of $L_z(\Omega)$, that is, Eq.(39) with $\hat{\rho _s}(T)/\rho =1$.

\begin{figure}
\includegraphics [scale=0.4]{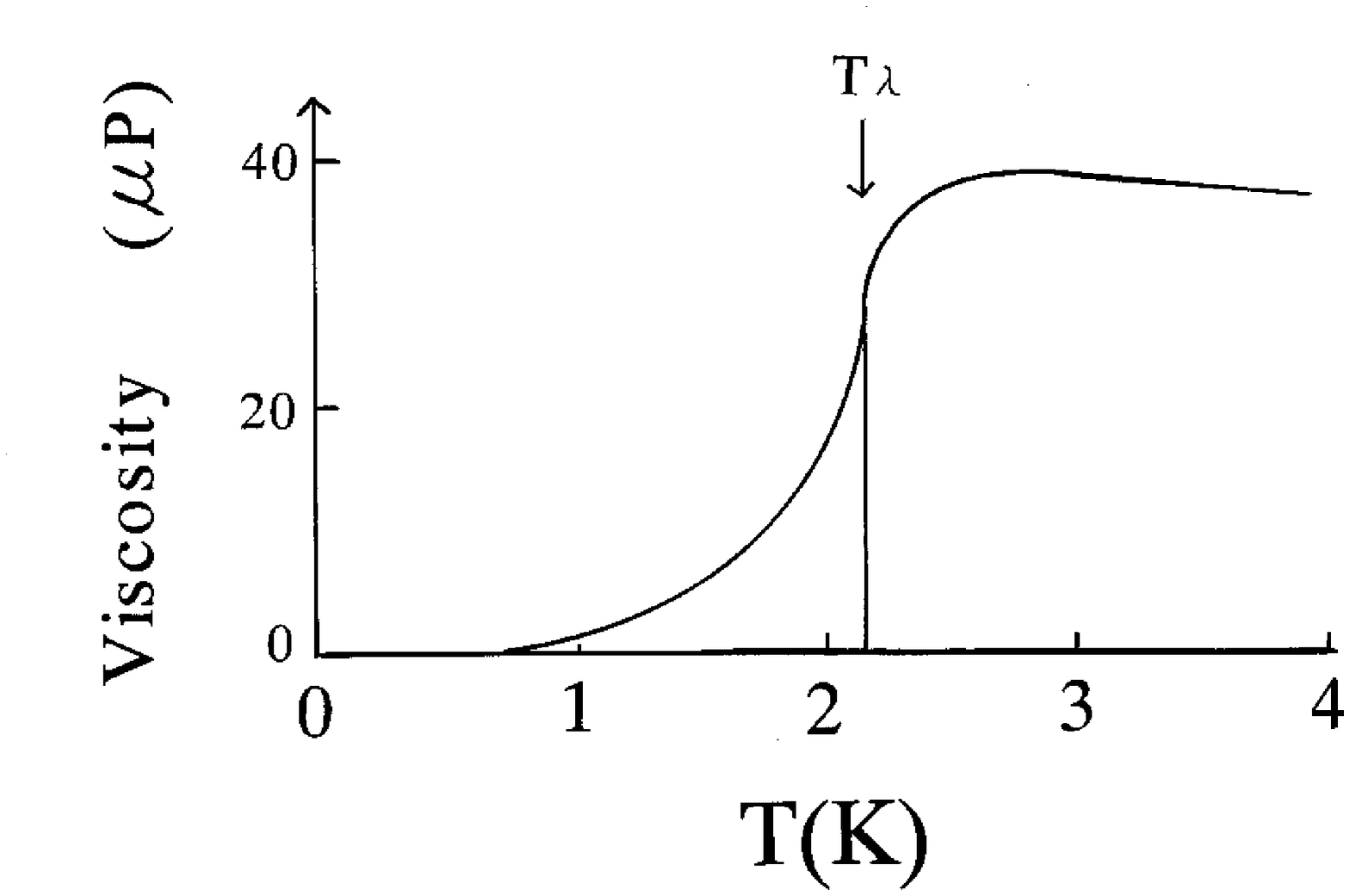}
\caption{\label{fig:epsart}
The temperature dependence of the shear viscosity of  a liquid 
  helium 4. (The data at $T>T_{\lambda}$ is taken from Fig.3 of Ref.\protect \cite {tay}.
   The data at $T<T_{\lambda}$ that strongly depends on the experimental method is written schematically.).}
\end{figure}

\section{ Discussion}

  \subsection{Shear viscosity at the vicinity of $T_{\lambda}$} 
 In a classical liquid, with decreasing 
temperature, the shear viscosity gradually increases. 
The shear viscosity of a liquid is inversely proportional to the rate of
a process in which a hole in a liquid propagates from one point to 
another over the energy barriers. With decreasing temperature, this rate 
decreases, thus increasing the viscosity. 
 In a liquid helium 4, however, it reaches a maximum value at 
2.8K, and begins to reduce its value, finally dropping at the $\lambda$ 
point (2.17K).  Figure.5 illustrates this temperature dependence taken 
from the data at $T>T_{\lambda}$ in \cite{tay}. 
In view of some precursory phenomena observed in a liquid helium 4, it is natural 
to explain it in terms of the thermal fluctuations giving rise to 
the short-lived wave function. 

 From the viewpoint of this paper, however, one must point out  
 two other  aspects of this phenomenon. 
 (1) Under the strong influence of Bose statistics just above 
$T_{\lambda}$, the large but not yet macroscopic long-lived coherent wave function 
must affect the response of the system. One must have a microscopic 
 theory describing the influence of superfluidity on the coefficient of 
 viscosity  $\eta$, irrespective of whether the decrease of $\eta$ is caused by the fluctuations 
 or by the long-lived wave functions.  For such a theory, (a) one 
 must apply the linear-response theory not to the mechanical, but to the 
 thermal perturbation \cite{kad}. The formulation of the latter 
 perturbation includes more subtle points 
 than that of the former one owing to the thermal dissipation. (b) The 
 mechanism of shear viscosity in a liquid, which has similarities with the 
 motion of dislocations in a solid, is considerably different from that 
 of shear viscosity in a gas. Hence, the 
 quasiparticle approximation,  which is often used for 
 the excitations in a liquid helium 4, is questionable for the shear 
 viscosity in a liquid. The microscopic theory of the 
influence of superfluidity on $\eta$ is a future problem.

(2) The viscosity appears in the Navier-Stokes equation as $\eta \Delta \mbox{\boldmath 
$v$}$, which has a form of $-\eta \cdot rot(rot \mbox{\boldmath 
$v$})$ in  an incompressible fluid.  
The decrease of the viscosity comes from either the decrease of $\eta$ or 
that of $rot \mbox{\boldmath $v$}$ \cite {lod}. The former is owing to the microscopic change of the system, 
whereas the latter arises from the macroscopic transformation of the velocity 
field $\mbox{\boldmath $v$}(\mbox{\boldmath $r$})$ in a container. 
 At $T<T_{\lambda}$, the temperature dependence of the shear viscosity strongly depends on the 
experimental methods, such as (1) a Poiseuille flow method (a lower 
horizontal line at $T<T_{\lambda}$ in Fig.5), (2) an oscillating disc viscometer 
method (an upper curve at $T<T_{\lambda}$) and (3) a rotation viscometer method.
Since the microscopic mechanism determining $\eta$ is not sensitive to 
the types of the macroscopic measurements, this fact suggests that the shear 
viscosity at $T<T_{\lambda}$ is strongly dependent 
on the macroscopic change of $rot \mbox{\boldmath $v$}$ induced by the measurements.  
Above $T_{\lambda}$, however, since the coherent wave 
function has not yet grown to a macroscopic size, 
 it is natural to attribute the decrease of viscosity at first to 
 that of $\eta$. Just above $T_{\lambda}$, however, we have another 
 possibility in $-\eta \cdot rot(rot \mbox{\boldmath $v$})$ that the 
emergence of the intermediate-sized coherent wave functions contributes 
to the decrease of $rot \mbox{\boldmath $v$}$, because they are irrotational.
 This means that the microscopic mechanism of a 
liquid affects not only the coefficient $\eta$, but also the macroscopic velocity field 
$\mbox{\boldmath $v$}(\mbox{\boldmath $r$})$. This is a peculiar case in fluid mechanics.
In fluid mechanics,  the velocity field $\mbox{\boldmath 
$v$}(\mbox{\boldmath $r$})$ is a solution of the equations of motion with  
given coefficients made of mechanical or thermodynamical constants of a liquid. In the 
system such as a liquid helium 4, one cannot clearly distinguish between the microscopic 
and macroscopic phenomena, and therefore, {\it this clear separation of coefficient 
and solution of the equation is not obvious \/} in contrast with a classical 
liquid.  One must make a different approach that will examine the 
foundation of the two-fluid mechanics at the vicinity of $T_{\lambda}$. 

 \subsection{Various manifestations of superfluidity} 
  Superfluidity is a complex of phenomena, and therefore  
 has some different manifestations, such as (1) persistent 
current without friction, (2) the Hess-Fairbank effect, (3) quantized 
circulation, (4) almost no friction on moving objects in the system below 
the critical velocity, (5) peculiar collective excitations and (6) the Josephson 
effect. The conventional thermodynamical definition of superfluid density 
$\rho _s(T)$ has been proved to be useful for describing various mechanical 
manifestations of superfluidity. The result of this paper implies 
that {\it the superfluid density in the  mechanical phenomena does not 
always agree with the thermodynamical 
$\rho _s(T)$, \/} and that the interplay between Bose statistics and the repulsive interaction 
sometimes require us to define {\it the mechanical superfluid density \/}
 $\hat{\rho _s}(T)$. This  $\hat{\rho _s}(T)$  may have some different
definitions, each of which is  specific to each manifestation of superfluidity, such
 as $\hat{\rho _s}(T)$ in Eq.(26) for the rotation. Hence, we must  
consider the existence or non-existence of $\hat{\rho _s}(T)$ in each 
manifestation on a case-by-case basis. Here we make a comment on three examples.

(A) The Meissner effect in the charged Bose system is a 
counterpart of the nonclassical rotational  behavior in the neutral 
Bose system \cite{noz}  \cite{bay}. (The enhanced diamagnetism above $T_{BEC}$ 
appears as the precursory phenomenon owing to thermal fluctuations \cite {gol}.)
The phenomenon like the decrease of $I_z$, not owing to fluctuations but requiring  a transformation 
extending to the whole system, occurs just above $T_{BEC}$ in the Meissner effect as well: With decreasing 
temperature, the charged Bose gas with short-range repulsion begins to 
exclude the magnetic field prior to the BEC \cite{koh}. In contrast with 
the enhanced diamagnetism, the exclusion of the 
applied magnetic field just above  $T_{BEC}$ implies that the large but not yet macroscopic 
wave functions cause a change extending to the whole system. This means that 
Bose statistics is essential for the Meissner effect, but the
existence of the macroscopic condensate is not the necessary condition. The same form of the 
mechanical superfluid density $\hat{\rho _s}(T)$ is useful for the 
Meissner effect as well. 

(B) The nonclassical rotational behavior above $T_{\lambda}$ will become 
more realistic in superfluidity of {\it small systems. \/} 
Recently, a helium 4 droplet consisting of about $10^4$ 
helium 4 atoms is found to show a sign of superfluidity. The infrared rotational spectrum of small 
molecules attached to the helium 4 droplet, such as oxygen carbon 
sulfide, shows a signal indicating a significant change of its moment of 
inertia around a molecular axis at $T<T_{\lambda}$, which suggests a transition occurring in the surrounding 
helium 4 environment  \cite{dro}. Although the experimental 
condition, such as the temperature or the rotational 
velocity, cannot be freely controlled until now as in the bulk helium 4, 
it will open a new probability of superfluidity of small systems. If the 
same experiment as in Sec.4 could be made for a liquid helium 4 
droplet, one would see a larger role of the intermediate-sized coherent wave function
$\hat{\rho _s}(T_1)/\rho$ in the rotational properties \cite {equ}.  
 When the number of atoms in a droplet is too small, however, the $\lambda$ 
 transition will become obscure. Hence, there may be an optimum size 
 of the system for detecting the nonclassical rotational behavior just above $T_{\lambda}$.

(C) Solid helium 4 has been termed a quantum crystal. Recently, an abrupt 
drop in the moment of inertia was found in the tortional oscillation measurements
 on solid helium 4 confined in a porous 
media \cite{cha} and on a bulk solid helium 4 \cite{kim}.  This discovery leads 
us to reconsider the definition of superfluidity and that of solids \cite {leg}. 
The fundamental feature of crystals is their periodicity in density; 
that is, diagonal long-range order (DLRO). One has to face a serious question whether 
 crystals remain stable while showing superfluidity that violates their periodicity. 
 Phenomenologically, this discovery shares the following point with the subject 
 of this paper: The nonclassical decrease of the moment of inertia 
 occurs even in the system in which the existence of the macroscopic Bose 
 condensate is not expected. There is, however, the following difference. In 
 a normal liquid helium 4 at the vicinity of $T_{\lambda}$, 
 helium 4 atoms actually exist, and the size of 
 their coherent wave function with zero momentum is a subject of the 
 problem, whereas in a solid helium 4, the existence of a hypothetical moving boson, a zero-point 
 vacancy, has not  been firmly established, although it is normally assumed to exist. 
 There are a number of problems to be clarified in a solid helium 4.

\appendix 
 \section{Comparison to the thermal fluctuation}
 Compared with the thermal-fluctuation theories, the formalism in Sec.3 
 has the following differences.
(1) In the fluctuation theory of a quantity $x$ in question, the 
correlation function of deviations $x-\bar {x}$ from a mean value $\bar {x}$ gives 
the susceptibility,  an average of which is taken with the Gauss distribution.  
Owing to the flatness of the minimum of the 
thermodynamic  potential near $T_c$, the susceptibility has $(1-T/T_c)$ 
dependence. (Alternatively, the introduction of Green's functions 
representing thermal fluctuations leads to the similar $(1-T/T_c)$ 
dependence.) On the other hand, the susceptibility 
in Sec.3 is given as the correlation function of total quantities $x$ as 
in Eq.(9). Its average is taken with the Bose-Einstein distribution. Hence, the 
result has no  $(1-T/T_c)$ dependence.
(2) Since thermal fluctuation is essentially a local phenomenon, one 
notes its influence already in the low-order terms of the perturbation 
expansion in which only a few particles participate. 
On the other hand, the change of the moment of inertia  $I_z$ requires 
a transformation extending to the whole liquid. Hence, the higher-order terms in which many 
particles participate plays a significant role. Only after the 
perturbation expansion is summed up to the infinite 
order as in Eq.(18), one notices the change of $I_z$ at a temperature in which Eq.(20) is satisfied. 

(3) Thermal fluctuations gradually increase as $T\rightarrow T_c$, 
finally diverging at $T_c$. For the change of the moment of inertia, 
however, no physical relevant quantity diverges at $T_c$,
and $I_z$ smoothly agrees with the conventional value of $I_z^{cl}(1-\rho 
_s/\rho)$ as in Eq.(26).

¥

%\end{references}



\begin{thebibliography}{99}
\bibitem{var} The variational theory at the vicinity of $T_{\lambda}$, 
        which assumes the expandibility of the thermodynamic potential in powers of the small order 
        parameter,  belongs to this category.
\bibitem{equ} In the equation of states 
         $N/V=g_{3/2}(e^{\beta\mu})+V^{-1} e^{\beta\mu}/(1-e^{\beta\mu})$ 
         where $g_{a}(x)=\sum_{n}x^n/n^{a}¥$, $V^{-1} e^{\beta\mu}/(1-e^{\beta\mu})$ has a 
         finite value at $V\rightarrow\infty$ limit only when 
         $e^{\beta\mu}/(1-e^{\beta\mu})$ diverges. Hence, this limit decomposes all 
         states into only two classes, the condensate  and the whole 
         of other states, and therefore each of other states loses its characteristics.
\bibitem{tre}  Whatever complicated approximation is made for 
            the susceptibility, this treatment enables us to derive the 
            superfluid density from it.
\bibitem{com} It is possible that  more complex diagrams 
            than the  bubble in $|G>$ may participate in 
            the particle exchange with the tensor $J_{\mu}J_{\nu}$ in 
            Fig.3(a).  Although it is 
            difficult to estimate an infinite sum of these diagrams, it adds only a small 
            correction to the whole feature of the phenomenon.
\bibitem{dif} For the differential rotation in astrophysics, such as that of rotating stars 
          or galaxies, the peripheral region   does not follow the rigid-body rotation, 
          because the driving force of rotation  is given at the center. 
          On the contrary, for the differential rotation of a superfluid in a bucket, 
          the central region does not follow the rigid-body rotation.
\end{thebibliography}
\end{document}